# Simple Coherent Polarization Manipulation Scheme for Generating High Power Radially Polarized Beam

P. B. Phua*
DSO National Laboratories,
20, Science Park Drive, S118230, Republic of Singapore,
Nanyang Technological University,
50, Nanyang Avenue, S 639798, Republic of Singapore

W. J. Lai
Temasek Laboratory
Nanyang Technological University,
50, Nanyang Avenue, S 639798, Republic of Singapore

* Corresponding Author: ppohboon@alum.mit.edu

**Abstract**:

We present a simple novel scheme that converts a Gaussian beam into an approximated radially polarized beam using coherent polarization manipulation together with Poynting walk-off in birefringent crystals. Our scheme alleviates the interferometric stability required by previous schemes that implemented this coherent mode summation using Mach-Zehnder-like interferometers. A symmetrical arrangement of two walk-off crystals with a half-wave plate, allows coherence control even when the laser has short temporal coherence length. We generated 14 watts of radially polarized beam from an Ytterbium fiber laser, only limited by the available fiber laser power.

Radially polarized beam has gained much interest recently [1-17], due to its ability to be focussed tighter than a diffraction-limited beam [1]. This can improve applications such as particle-trapping, optical data storage, laser machining and micro-lithography. In addition, the presence of high intense longitudinal electric field in the vicinity of the focal point of a laser beam can also enhance nonlinear effects for applications such as tip-enhanced Raman spectroscopy [2]. This longitudinal electric field can also be used for laser particle acceleration without a plasma wave. Most reported methods generate radially polarized beam by placing specially designed optical elements [4-10] inside the laser cavity. However, this introduces additional intra-cavity loss, and may make the optimal design of the laser difficult. Moreover, for high power fiber lasers, UV lasers used in lithography, semiconductor lasers, it is not always viable to add such optical elements inside the laser resonator. Thus, an external radially polarization conversion is an attractive and flexible alternative [1,11-16]. Previous reported external polarization conversion schemes based on a liquid crystal array [11] and diffractive phase element [12], tend to have low power handling while schemes that use segmented half wave-plates scheme [1] can only approximate the radially polarized beam. While the spirally varying retarder proposed recently in [14] has high laser power handling capability, it requires specialized fabrication technique for its spiral profile.

Radially polarized beam is a coherent summation of a horizontal polarized $TEM_{10}$ with a vertical polarized $TEM_{01}$ Hermite-Gaussian mode [6,13,15,16]. Previous interesting schemes [6,13,15,16] implemented this coherent summation of modes using Mach-Zehnder-like interferometric arrangements, in the effort of converting Gaussian beam into a radially polarized beam. Most recently, Ref. [17] proposed the coherent mode transformation in the reverse manner, from a radially polarized beam to an approximated Gaussian mode. The main limitation of these interferometric methods is they require interferometric stability. This may limit its practical usefulness. In addition, they also require specially designed optics such as spiral phase plate and binary diffractive optical element, which may not be widely available as they require special fabrication techniques.

In this paper, we propose a simple and stable coherent polarization manipulation scheme to convert a Gaussian mode to an approximated radially polarized beam. The scheme makes use of Poynting walk-off effect in birefringent crystals together with coherent polarization manipulation accomplished using standard off-the-shelf polarization wave-plates. In this scheme, we not only manipulate the polarizations of the various parts of the beam, we also maintain certain phase relationship between them, so that they interfere appropriately to generate an approximated radially polarized beam. For this reason, we call it a coherent polarization manipulation scheme. This manipulation, being carried out inside birefringent crystals instead of free-space interferometers, offers the compactness, mechanical stability and robustness demanded by practical applications. We have generated 14 Watts of approximated radially polarized beam from a Ytterbium (Yb) fiber laser. The demonstrated power is only limited by the available power from our fiber laser. However, since all optical components used in our scheme are standard off-the-shelf components that have power handling capability of more than kilowatts of laser power, the scheme will be a promising radial polarization converter for high power fiber laser.

Our simple scheme, in principle, requires only three optical components: a Poynting walk-off crystal, a thin half-wave plate and a $45^o$ quartz rotator, as shown in Figure 1. The laser beam cross-section after each stage of the scheme, and their associated polarizations are also shown in Figure 1. An input elliptical Gaussian beam with a $45^o$ linear polarization is being splitted by a Poynting walk-off crystal into two equal power beams with a beam separation that determined by the crystal's walk-off length. The polarizations of the top and bottom beam are orthogonal and their fields are controlled to be in phase by slight tilting the walk-off crystal. A thin half-wave plate (HWP), with slow axis of 45$^o$ with respect to the horizontal direction, is then used to rotate the polarization of the right half of both the top and bottom elliptical beams by 90$^o$. By tilting HWP about the vertical direction, as shown in Figure 1, we control the phase difference between the left and the right lobes of the beams to be 180$^o$ out of phase. A subsequent 45$^o$ optical activity quartz rotator then rotates the polarizations of the four lobes globally by 45$^o$. The generated 4-lobes beam, and their respective polarizations (as shown in Figure 1) form a set of approximated Hermite-Gaussian TEM$_{10}$ and TEM$_{01}$ modes in a 45$^o$ rotated frame. These TEM$_{10}$ and TEM$_{01}$ modes are of orthogonal polarizations and are coherently in-phase. As suggested by Kogelnik and Li back in 1966 [18], these modes can superimpose coherently in far-field to approximate the radially polarized beam.

We first use a continuous-wave, single longitudinal mode, 1064nm laser to demonstrate this scheme. This highly coherent laser source has power only up to a few milli-watts. With a spatial filter that allows ~75% power throughput, we generated a nearly radially polarized beam as shown in Figure 2, with a beam quality of $M^2$~2.2. Without a polarizer, the output was a doughnut-shaped light beam as shown in Figure 2(a). When the polarizer was inserted prior to the camera, two spots were clearly seen, and they rotated with the transmitting axis of the polarizer, as seen in Figure 2(b)-(e) which clearly illustrated the radially polarized profile. The mode conversion is stable and it lasts many hours to several days without any tweeking. This is expected since all polarization and phase manipulations occur inside the crystals and waveplates. Such stability is much better than that of free-space interferometric methods suggested in [6,13,15,16,17].

To scale up the power level ( >10 Watts) of the radially polarized light, we changed the laser source to a Ytterbium (Yb) doped fiber laser that can produce 15 Watts of polarized 1064 nm. This fiber laser has a broad bandwidth of 2 nm and our initial attempt to perform radial polarization conversion using the three-element scheme shown in Figure 1, was unsuccessful. This is because the coherent length of this laser is shorter than the optical path difference (i.e $(n_e - n_o) l$ ) traversed by the two orthogonally polarized beams in the walk-off crystal. Noted that $n_o$ and $n_e$ are the respective refractive indices of the ordinary and extra-ordinary polarization of the crystals. This inhibits any coherent mode superposition for the generating radially polaried light. To circumvent this issue, we used two identical birefringent crystals (Walk-Off Crystal 1 and Walk-Off Crystal 2) with a half-wave plate (HWP2) placed between the two crystals, as shown in Figure 3, to perform the beam separation. The crystallographic optics axis of Walk-Off Crystal 1 is $45^o$ with respect to the laser propagation direction while that of Walk-Off Crystal 2 is $135^o$. Each crystal has length $l$ and has a Poynting walk-off distance of $d_w$ for their extra-ordinary polarization. Figure 3(b) – 3(h) show the laser beam cross-sections and their associated polarizations after each stage of the scheme. Upon entering the Walk-Off Crystal 1, the extra-ordinary polarization component (i.e. vertical polarization) walk-off vertically upward with respect to the ordinary polarization component (i.e. horizontal polarization). With the slow axis of HWP2 being $45^o$ with respect to the horizontal direction, the top elliptical beam after passing though HWP2, becomes the ordinary polarization while the bottom becomes the extra-ordinary polarization of Walk-Off Crystal 2. The bottom elliptical beam therefore walks a distance of $d_w$ vertically downward after passing through Walk-Off Crystal 2. The total separation between the two elliptical beams after passing through the two crystals is $2d_w$. It is worthwhile to note that, with such two-crystal arrangement, both top and bottom elliptical beam have traversed the same optical length of $(n_o + n_e) l$. Thus, constant coherent phase relationship can be maintained between the top and bottom elliptical beam even when the laser has short temporal coherence length. Thus this allows coherent superposition of modes even for the broadband Yb fiber laser.

In addition, in this particular experiment, instead of rotating the Walk-off crystal, as shown in Figure 1, to maintain same phase for the top and bottom elliptical beams, we use a polarization controller comprising of a half-wave plate (HWP1) followed by a quarter-wave plate (QWP) to accomplish the same effect. The input polarization of the elliptical beam, in this case, is vertically linear. The slow axis of QWP is fixed at 45° with respect to the horizontal direction while the slow axis of the half-wave plate is adjustable. By rotating the slow axis of HWP1, we adjust the ellipticity of the 45° oriented input polarization that enters Walk-Off crystal 1, so that the top and bottom elliptical beam exiting from Walk-Off Crystal 2, have equal power and phase. This accomplishes the same effect as rotating the Walk-off crystal in Figure 1. In this experiment, the 1-micron Ytterbium (Yb) doped fiber laser beam was shaped using a pair of cylindrical lenses into an elliptical Gaussian profile of $\omega_x$ of 2.7 mm and $\omega_y$ of 1.35 mm. The two crystals used were identical Alpha-BBO crystals (45°-cut) of length 11.4 mm. Each crystal contributes a walk-off distance of 1 mm and their orientations were accordingly to Figure 3. For HWP3, we used a 60.8 microns thick crystalline quartz wave-plate which is a true zero-order half-wave plate, and its slow axis is 45° with respect to the horizontal direction. The small thickness of HWP3 gives robust angular tolerance in controlling the phase difference between the left and right lobes of the beams.

We obtained 14 watts of average power of approximated radially polarized beam, only limited by the maximum power available from our fiber laser. Without any thermal management for the walk-off crystals and polarization wave-plates, we did not observe any thermal fluctuation at this operating power. This endorses the fact that the alpha-BBO crystals and the crystalline quartz wave-plates have negligible laser absorption and high power handling capability. Thus, we are optimistic that the scheme has good potential to produce more than hundreds of watts of radially polarized light. The approximated radially polarized beam, has the measured beam quality was $M_x^2 \sim 3.6$ and $M_y^2 \sim 2.1$. Figure 4 shows the radially polarized beam (observed with and without polarizer) after a spatial filtering of power throughput of 73%. The measured beam quality after filtering was $M_x^2 \sim 2.43$ and $M_y^2 \sim 2.28$.

In conclusion, we have presented a novel scheme that starts with a Gaussian mode, and using Poynting walk-off effect in birefringent crystals together with coherent polarization manipulation, the scheme converts it to an approximated set of orthogonally polarized and coherently in-phase $TEM_{10}$ and $TEM_{01}$ modes. These modes superimpose at far field to generate a near radially polarized beam. Since the manipulation is accomplished inside the crystals, it alleviates the high interferometric sensitivity of previous interferometric arrangements and offers a much compact, stable, and robust scheme to convert a Gaussian mode into a radially polarized beam. We demonstrated 14 watts of radially polarized beam but the scheme has potential to produce much higher power of radially polarized beam. In addition, we have shown a symmetrical arrangement of two walk-off crystals with a half-wave plate so that the temporal coherence length of the laser will not pose a problem to the coherent manipulation of the beam. The authors would like to acknowledge Lim Yuan Liang, Lai Kin Seng, Tan Beng Sing for their helpful discussion, and Teo Kien Boon for his support and encouragement in this work.

**Figure Captions**

Figure 1 :   Schematic of our simple coherent polarization manipulation scheme which comprised of a walk-off crystal, a thin Half-Wave Plate (HWP) and a 45$^o$ quartz rotator. Their laser beam cross-sections and their associated polarizations after each component are also shown.

Figure 2 :   Radially polarized beam converted from a highly coherent, single longitudinal mode 1064 nm laser: (a) observed without polarizer, and observed with polarizer whose transmitting axis is (b) 0$^o$, (c) 90$^o$ (d) 45$^o$, and (e) 135$^o$.

Figure 3 :   Schematic of our coherent polarization manipulation scheme for low coherent broadband Yb fiber laser. Laser beam cross-sections and their associated polarizations of (b) input elliptical Gaussian beam, (c) after QWP, (d) after Walk-Off Crystal 1, (e) after HWP2, (f) after Walk-Off Crystal 2, (g) after HWP3, (h) after quartz rotator.

Figure 4 :   Radially polarized beam converted from a low coherent, broadband 1064 nm Yb fiber laser: (a) observed without polarizer, and observed with polarizer whose transmitting axis is (b) 0$^o$, (c) 90$^o$ (d) 45$^o$, and (e) 135$^o$.

.

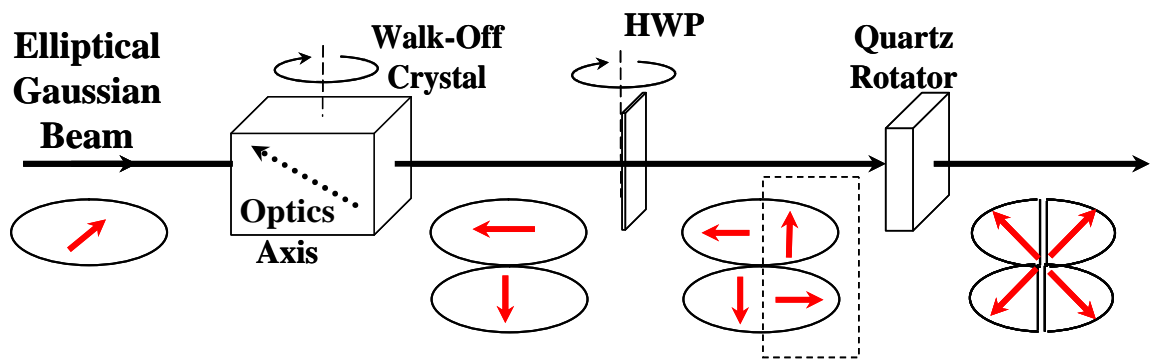

**Figure 1**

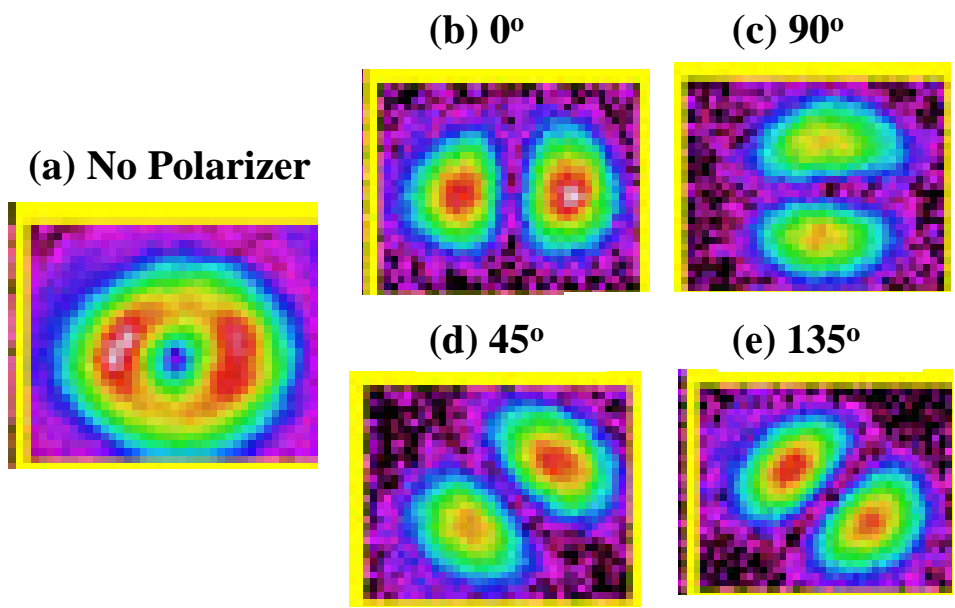

**Figure 2**

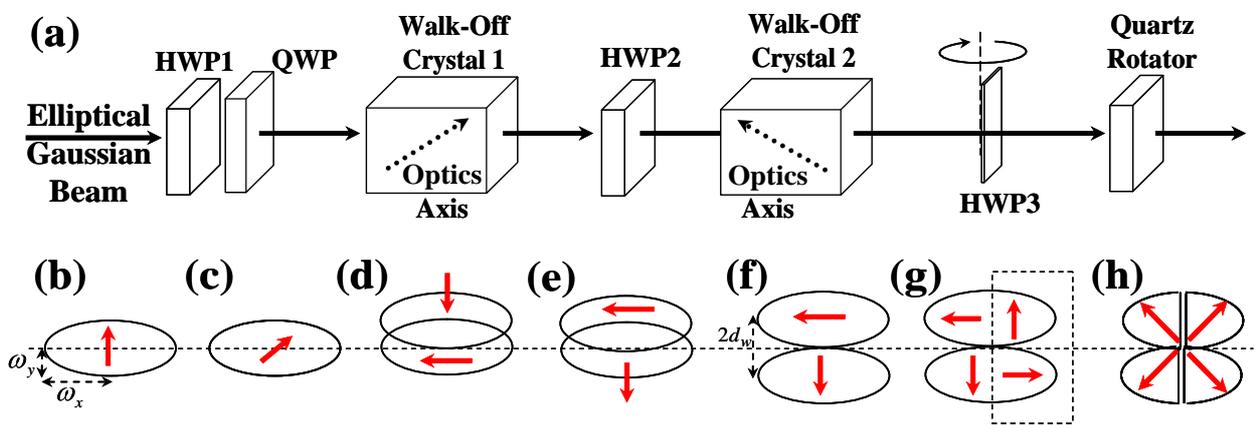

**Figure 3**

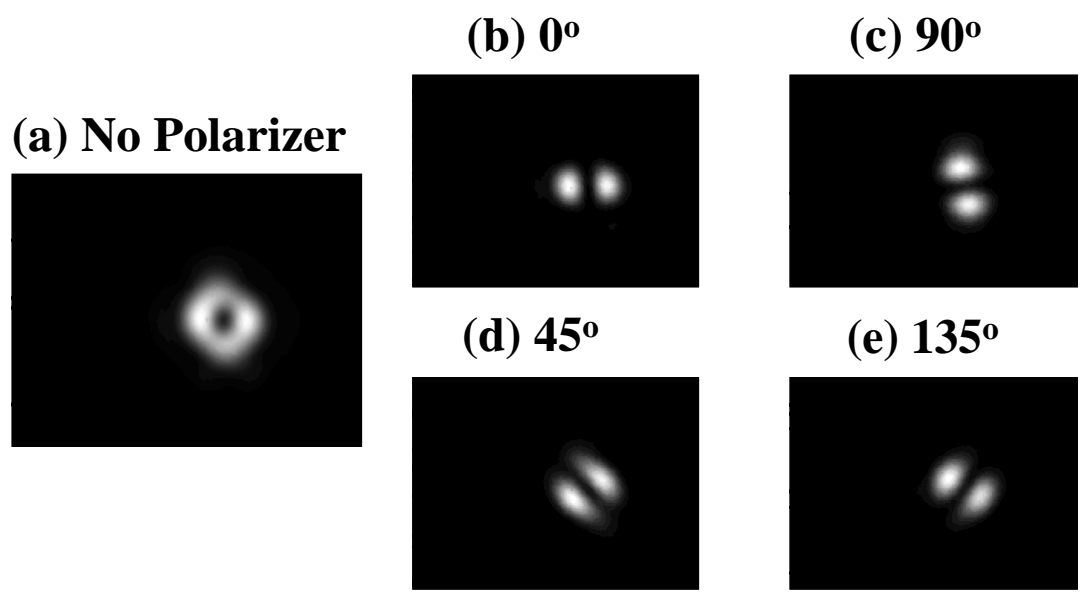

**Figure 4**